\documentclass[a4paper]{article}
\usepackage{fancyvrb}

\fvset{fontsize=\scriptsize}
\title{Geometric Memory Management}
\author{Wouter Kuijper}
\date{\today\\\ \\\ \\DRAFT VERSION\\}
\begin{document}
\maketitle \abstract{
  In this report we discuss the concepts of geometric memory alignment,
  geometric memory allocation and geometric memory mapping.
  We introduce block trees as an efficient data structure for
  representing geometrically aligned block allocation states.
  We introduce niche maps as an efficient means to find the right
  place to allocate a chunk of a given size whilst maintaining good
  packing and avoiding fragmentation.
  We introduce ledging as a process to deal with chunk sizes that are
  not a power of two.
  We discuss using block trees for memory mapping in the context of
  virtual memory management.
  We show how ledging can also be applied at the level of virtual
  memory in order to create fixed size virtual memory spaces.
  Finally we discuss implementation of both geometric memory
  allocation as well as geometric memory mapping in hardware.
}
\section{Geometric Memory Allocation}\label{sec:introduction}
Traditional memory allocation schemes, with the exception of some
clean implementations of the buddy block allocation
scheme~\cite{knuth1972art}, typically align slots up to eight,
sometimes sixteen bytes.

This pervasive design choice leads to what we will refer to as the
\emph{misaligned slots problem}, which in turn leads to a myriad of
fragmentation problems down the road.

The misaligned slot problem occurs when two consecutive slots of
size $2^n$ are being coalesced into a bigger slot of size $2^{n+1}$
\emph{on an address that is not a multiple of} $2^{n+1}$.

A simple way to avoid these problems altogether is to: \emph{align
  slots to their own size}. If we follow the latter alignment
principle we end up with a \emph{geometric allocation scheme}.

To see why the misaligned slot problem leads to fragmentation, and to
see how a geometric allocation scheme remedies this, consider the
following example.


Say for sake of simplicity we have a memory of 16 bytes, and we are
aligning to words of 2 bytes each. We initially consider the case of
just two \emph{size classes}. We have {large} slots of 2 words (4
bytes) in size, and \emph{small} slots of 1 word (2 bytes) in
size. Now say we have two allocated large slots at the start and at
the end of the memory, and four allocated small slots in between:
\begin{Verbatim}
   Step 1:    |xxxxxxxxxxxxxxx|xxxxxxx|xxxxxxx|xxxxxxx|xxxxxxx|xxxxxxxxxxxxxxx|
              | 0 | 1 | 2 | 3 | 4 | 5 | 6 | 7 | 8 | 9 | A | B | C | D | E | F |
\end{Verbatim}
Next we consider what happens when two of the smaller slots in the
center get freed:
\begin{Verbatim}
   Step 2:    |xxxxxxxxxxxxxxx|xxxxxxx|-------|-------|xxxxxxx|xxxxxxxxxxxxxxx|
              | 0 | 1 | 2 | 3 | 4 | 5 | 6 | 7 | 8 | 9 | A | B | C | D | E | F |
\end{Verbatim}
Most non--geometric allocators would coalesce these two freed, small
slots into one large slot:
\begin{Verbatim}
   Step 3:    |xxxxxxxxxxxxxxx|xxxxxxx|---------------|xxxxxxx|xxxxxxxxxxxxxxx|
              | 0 | 1 | 2 | 3 | 4 | 5 | 6 | 7 | 8 | 9 | A | B | C | D | E | F |
\end{Verbatim}
Next we consider what happens when we subsequently receive a request
for a large slot and we allocate it in the center:
\begin{Verbatim}
   Step 4:    |xxxxxxxxxxxxxxx|xxxxxxx|xxxxxxxxxxxxxxx|xxxxxxx|xxxxxxxxxxxxxxx|
              | 0 | 1 | 2 | 3 | 4 | 5 | 6 | 7 | 8 | 9 | A | B | C | D | E | F |
\end{Verbatim}
We end up with a misaligned large slot adorned by two smaller
slots. The center slot is misaligned because it does not occur at an
address that is a multiple of its own size, i.e.: 6 is not a multiple
of 4.

Next we consider the resulting state after the two remaining smaller
slots get freed:
\begin{Verbatim}
   Step 5:    |xxxxxxxxxxxxxxx|-------|xxxxxxxxxxxxxxx|-------|xxxxxxxxxxxxxxx|
              | 0 | 1 | 2 | 3 | 4 | 5 | 6 | 7 | 8 | 9 | A | B | C | D | E | F |
\end{Verbatim}
We have reached a fragmented state. Both of the small, free slots have
become \emph{squashed} between large, allocated slots. Because the
free slots are separated from each other we have no chance to coalesce
them.

One might say this is just a temporary situation; after all we need
merely wait for the center large slot to become freed and thereby the
small slots to become \emph{UN squashed}. However profiling most memory
loads shows that the residence time of a slot is proportional to its
size. Hence it is not hard to see this problem of fragmentation
becomes systemic.

Let us revisit Step 2, now using a geometric scheme. In order not to
confuse terminology, in the geometric case we speak of \emph{blocks}
instead of slots. In this paper, blocks are \emph{almost} the same
concept as slots yet they are subtly different in the sense that they
satisfy the geometric alignment criterion, that is, they start at an
address that is a multiple of their size.
\begin{Verbatim}
   Step 2:    |xxxxxxxxxxxxxxx|xxxxxxx|-------|-------|xxxxxxx|xxxxxxxxxxxxxxx|
              | 0 | 1 | 2 | 3 | 4 | 5 | 6 | 7 | 8 | 9 | A | B | C | D | E | F |
\end{Verbatim}
Because address 6 is not a multiple of 4 we will \emph{not} coalesce
into a large block. This is the central trade--off that a geometric
memory allocator makes: \emph{it occasionally foregoes capacity in
  larger size classes for the sake of preventing fragmentation}.

If next the rightmost of the small blocks frees up we end up in the
following situation:
\begin{Verbatim}
   Step 5.1:  |xxxxxxxxxxxxxxx|xxxxxxx|-------|-------|-------|xxxxxxxxxxxxxxx|
              | 0 | 1 | 2 | 3 | 4 | 5 | 6 | 7 | 8 | 9 | A | B | C | D | E | F |
\end{Verbatim}
We see that address 8 is a multiple of 4 and therefore we may coalesce
the rightmost two smaller blocks into one large block:
\begin{Verbatim}
   Step 5.2:  |xxxxxxxxxxxxxxx|xxxxxxx|-------|---------------|xxxxxxxxxxxxxxx|
              | 0 | 1 | 2 | 3 | 4 | 5 | 6 | 7 | 8 | 9 | A | B | C | D | E | F |
\end{Verbatim}
If next also the leftmost of the small blocks frees up, we see that we
coalesce again:
\begin{Verbatim}
   Step 5.2:  |xxxxxxxxxxxxxxx|---------------|---------------|xxxxxxxxxxxxxxx|
              | 0 | 1 | 2 | 3 | 4 | 5 | 6 | 7 | 8 | 9 | A | B | C | D | E | F |
\end{Verbatim}
This scheme scales in the obvious way. In particular if the left large block
frees up we may introduce a \emph{big} block:
\begin{Verbatim}
   Step 5.2:  |-------------------------------|---------------|xxxxxxxxxxxxxxx|
              | 0 | 1 | 2 | 3 | 4 | 5 | 6 | 7 | 8 | 9 | A | B | C | D | E | F |
\end{Verbatim}
And if now the right large block frees up we may introduce a
\emph{huge} block spanning the entire memory:
\begin{Verbatim}
   Step 5.2:  |---------------------------------------------------------------|
              | 0 | 1 | 2 | 3 | 4 | 5 | 6 | 7 | 8 | 9 | A | B | C | D | E | F |
\end{Verbatim}

With this example we have already seen the essence of geometric memory
allocation, which is really quite simple.
The rest of this report section addresses the following implementation
challenges:
\begin{enumerate}
\item Dealing with chunk sizes that are not a power of two.
\item Efficiently representing allocated blocks in memory.
\item Finding the right block for a requested allocation.
\end{enumerate}
For solving the first problem we introduce a process called
\emph{ledging} in Section~\ref{sec:ledging}. For solving the second
problem we introduce a data structure called a \emph{block tree} in
Section~\ref{sec:block_trees}. For solving the third problem we extend
block trees with \emph{niche maps} in Section~\ref{sec:niche_maps}.

\subsection{Ledging}\label{sec:ledging}

If a geometric allocator would be limited to blocks of size $2^n$ it
would be severely limited in utility. There are only a small number of
cases where applications request buffers that are guaranteed to be of
exponential size. Most of them involve dynamic arrays, hash tables and
the like. This is assuming we do not use a mapping layer as explained
in the next section in which case it is indeed sufficient for the
underlying allocation layer to only provide blocks sized to powers of
two.

For other use cases, however, the allocator must be able to accommodate
arbitrarily sized chunks. Of course we could stuff any given chunk in
its smallest enclosing power of two. However, in the worst case, a
chunk size of $2^n+1$, this would give an asymptotic overhead of 100\%
as $n$ grows larger. Clearly that would not be acceptable.

Therefore, rather than taking the smallest enclosing power of two, we
start by allocating the largest power of two that is smaller than the
chunk size and we work from there, making up for the excess using a
partial geometric series. The important property of the geometric
series is that this can be done \emph{exponentially quickly}. In
particular this will incur only constant, worst--case, overhead for
a given fixed pointer--width.

Since we are able to represent blocks to arbitrary precision
(byte--level) we are, in principle, able to allocate chunks of
arbitrary size to byte precision and allocate the rest of the space to
other chunks. In practice we will always work with a minimal chunk
size, since byte sized chunks simply do not constitute a useful size
class. For the purpose of exposition, however, we will continue to
show how blocks can be represented to the level of individual bytes.

\subsection{Block Trees}\label{sec:block_trees}

As an example of a block tree consider:
\begin{Verbatim}
 Level 4:     .---------------------------------------------------------------.
              :                                                               :
 Level 3:     .-------------------------------.-------------------------------.
              :                               :                               :
 Level 2:     .---------------.---------------.---------------.---------------.
              :               :               :               :               :
 Level 1:     .- - - -.- - - -.-------.-------.- - - -.- - - -.- - - - - - - -.
              :       |       :       :       :       :       :               :
 Level 0:     .- -.- -.- -.- -.- -.- -.- -.- -.- -.- -.- -.- -.- -.- -.- -.- -.
              :   :   :   :   :   :   :   :   :   :   :   :   :   :   :   :   :
              |xxxxxxxxxxxxxxx|xxxxxxx|-------|---------------|xxxxxxxxxxxxxxx|
              | 0 | 1 | 2 | 3 | 4 | 5 | 6 | 7 | 8 | 9 | A | B | C | D | E | F |
\end{Verbatim}

As an example of a \emph{sparse} block tree consider:
\begin{Verbatim}
 Level 4:     .---------------------------------------------------------------.
              :                                                               :
 Level 3:     .-------------------------------.-------------------------------.
              :                               :                               :
 Level 2:     .---------------.---------------.- - - - - - - -.---------------.
              :               :               :               :               :
 Level 1:     :               .-------.- - - -.               :               :
              :               :       :       :               :               :
 Level 0:     :               :       :       :               :               :
              :               :       :       :               :               :
              |xxxxxxxxxxxxxxx|xxxxxxx|-------|---------------|xxxxxxxxxxxxxxx|
              | 0 | 1 | 2 | 3 | 4 | 5 | 6 | 7 | 8 | 9 | A | B | C | D | E | F |
\end{Verbatim}
In the sparse block tree, the dashed lines indicate missing siblings,
these correspond exactly to the free blocks called \emph{niches}.

\subsection{Niche Maps}\label{sec:niche_maps}

As an example of a sparse block tree with niche maps consider:
\begin{Verbatim}
 Level 4:     .----------------------------[0110]-----------------------------.
              :                                                               :
 Level 3:     .------------[010]--------------.------------[100]--------------.
              :                               :                               :
 Level 2:     .-----[00]------.-----[10]------.- - - - - - - -.-----[00]------.
              :               :               :               :               :
 Level 1:     :               .--[0]--.- - - -.               :               :
              :               :       :       :               :               :
 Level 0:     :               :       :       :               :               :
              :               :       :       :               :               :
              |xxxxxxxxxxxxxxx|xxxxxxx|-------|---------------|xxxxxxxxxxxxxxx|
              | 0 | 1 | 2 | 3 | 4 | 5 | 6 | 7 | 8 | 9 | A | B | C | D | E | F |
\end{Verbatim}

A niche map is a vector of bounded precision, truncating counters that
denote lower bounds on the number of niches present in the levels
\emph{under} the level of the current node.

For example, the single node on level 1 is annotated with a single bit
indicating the presence or absence of level 0 niche(s), in this case
the value is [0] because the entire level 1 block is allocated (and
hence there are no free level 0 niches below it). As another example
the second node on level 2 is annotated with two counters indicating
the presence of level 1 and level 0 niches respectively, in this case
the value is [1,0] since there are level 1 niches but no level 0
niches. Finally, the root is annotated with four counters indicating
the presence of a level 3, 2, 1 and 0 niche respectively, in this case
the value is [0,1,1,0] because, in the entire tree, there are only
niches in level 2 and level 1.

Niche maps combine through simple truncating addition operations. To
get the niche map for a parent simply take the point wise truncated sum
of the niche maps of its children, and prepend a 1 iff there is a
missing child which implies the presence of a niche (as indicated by
the dashed lines).

As a special case: if we lower the precision of all counters to a
single bit, niche maps become bitmaps, and combining them reduces to a
simple bitwise OR operation.

\subsection{Best Fit Placement Strategy}

Using niche maps to guide us, we now have a very simple algorithm to
find a niche for a given chunk--size.

First, we look at the niche map for the root node, this gives us a best
fit niche (smallest niche that is larger than the requested
chunk--size). Or, in case there is no niche that is large enough, we know
that we cannot allocate the chunk and we give an out--of--memory error
back.

Next, we traverse the tree looking for the niche using the niche maps
to guide us going left or right at each binary junction. We are
guaranteed to eventually land on some niche of the desired size. We
can then proceed to allocate the chunk using the ledging process.

Finally we traverse back over the tree retrace--ing our steps all the
way back up to the root, updating the niche maps a we go.

Note that the above algorithm is non--deterministic: we can have ties
where it is both possible to go left or right. This provides a lot of
engineering freedom for coming up with particular placement strategies
which is useful for obtaining certain wear leveling or wear focusing
characteristics.

\section{Geometric Virtual Memory}

In the previous section we addressed geometric memory allocation in
the direct sense. In this section we describe how to virtualize memory
using a geometric approach.

Virtual memory management decouples application programs from the physical
memory. There is a host of reasons why this is important, we will not
treat those in this report.

Virtual memory management, in most current implementations, is done
through some form of paging. That is: whenever a read or a write
occurs somewhere in the virtual memory space a physical memory page is
allocated just in time to ``back'' this location.

The problem with constant page sizes is that this scheme does not
scale very elastically. If we want to have multiple, very lightweight,
virtualized address spaces, and if we want to be able to ``tolerate
holes'' in the real memory efficiently, it is desirable to have a more
lightweight, flexible scheme. Using the geometric alignment principle
it is not very hard to come up with an alternative, geometric virtual
memory management solution that provides just these characteristics.

The main idea is that, instead of constant page sizes, we allow blocks
to be remapped at exponentially differing scales using, again, block trees.

We can then have various population strategies for backing accesses that
the application makes to the virtual address space.
In particular it is possible to emulate a fixed size paging scheme, it
is also possible to provide a geometric sequence of blocks that grows
exponentially, doubling in size whenever an ``adjacent'' (also the
notion of adjacency here should be understood in an exponentially
widening sense, for details confer to the example below) memory
location is written to or read from.

The latter scheme has the advantage that it starts out very small and
scales up rapidly avoiding quadratic scaling overhead typically seen
in linear scaling approaches.

Another approach would be to have dedicated virtual spaces even for
allocations of fixed size buffers. This has the potential to improve
security by memory isolation. The extreme case would be to have an
object system where each and every object in memory receives its own
virtual space. This is possible with block trees by virtue of them
being significantly more lightweight than table based approaches.

\subsection{Using Block Trees for Memory Mapping}

The resulting data structure is again a sparse annotated block tree,
this time representing the virtual address space. The nodes in this
\emph{virtual block tree} are then annotated with offsets into the
real memory that backs these blocks (we will not show these blocks
below just indicate with a 'b' that a block is backed by \emph{some}
chunk of allocated real memory). It is possible, although not
necessary, to use also a geometric allocation scheme for managing the
real memory that backs the virtual memory.

As an example of this consider the following. Say we have again for
purposes of exposition, a virtual memory space consisting of just 16
addresses. Initially this space is pristine in the sense that no
single address has been touched and therefore no real memory has been
allocated to back it. The block three consist of a single niche:

\begin{Verbatim}
 Level 4:     .- - - - - - - - - - - - - - - - - - - - - - - - - - - - - - - -.
              :                                                               :
              |---------------------------------------------------------------|
              | 0 | 1 | 2 | 3 | 4 | 5 | 6 | 7 | 8 | 9 | A | B | C | D | E | F |
\end{Verbatim}

Next assume, again for purpose of exposition, that rather than
accessing the lowest or highest addresses in the virtual space first,
we have an application that accesses the virtual memory space dead
center first. So let us assume the application accesses memory
location 8. Because we do not know yet whether this memory access will
expand into a large buffer or remain limited to only a handful of
bytes it would be heavy handed to immediately allocate a large portion
of real memory to this first access. Typically some \emph{minimal}
block size will apply. However for the purpose of this example we will
use a minimal block size of just a single byte. This results in the
following virtual block tree:

\begin{Verbatim}
 Level 4:     .-------------------------------0-------------------------------.
              :                                                               :
 Level 3:     .- - - - - - - - - - - - - - - -.----------------0--------------.
              :                               :                               :
 Level 2:     :                               .--------0------.- - - - - - - -.
              :                               :               :               :
 Level 1:     :                               .---0---.- - - -.               :
              :                               :       :       :               :
 Level 0:     :                               .-1-.- -.       :               : 
              :                               :   :   :       :               :
              |-------------------------------|bbb|---|-------|---------------|
              | 0 | 1 | 2 | 3 | 4 | 5 | 6 | 7 | 8 | 9 | A | B | C | D | E | F |
\end{Verbatim}

Note that the size of this sparse block tree is still linear in the
pointer width. Because we need not determine our own allocations (that
is entirely up to the application that is \emph{using} the virtualized
memory space, not to the virtual memory manager), rather than keeping
a full niche map we have to keep only a single bit of information per
tree node and that is the \emph{full} bit. The full bit is set iff the
block is \emph{fully backed} by physical memory. Note that instead of
'x' for allocated memory we write 'b' for \emph{backed} virtual memory
locations. Now let us consider what happens if the following access
occurs in location 9:

\begin{Verbatim}
 Level 4:     .-------------------------------0-------------------------------.
              :                                                               :
 Level 3:     .- - - - - - - - - - - - - - - -.----------------0--------------.
              :                               :                               :
 Level 2:     :                               .--------0------.- - - - - - - -.
              :                               :               :               :
 Level 1:     :                               .---1---.- - - -.               :
              :                               :       :       :               :
 Level 0:     :                               .-1-.-1-.       :               : 
              :                               :   :   :       :               :
              |-------------------------------|bbb|bbb|-------|---------------|
              | 0 | 1 | 2 | 3 | 4 | 5 | 6 | 7 | 8 | 9 | A | B | C | D | E | F |
\end{Verbatim}

As expected we have now backed also location 9. As a result the level
1 block from 8 to 9 is also fully backed. Now let us consider what
happens if the following access occurs in location B:

\begin{Verbatim}
 Level 4:     .-------------------------------0-------------------------------.
              :                                                               :
 Level 3:     .- - - - - - - - - - - - - - - -.----------------0--------------.
              :                               :                               :
 Level 2:     :                               .--------1------.- - - - - - - -.
              :                               :               :               :
 Level 1:     :                               .---1---.---1---.               :
              :                               :       :       :               :
 Level 0:     :                               .-1-.-1-.       :               : 
              :                               :   :   :       :               :
              |-------------------------------|bbb|bbb|bbbbbbb|---------------|
              | 0 | 1 | 2 | 3 | 4 | 5 | 6 | 7 | 8 | 9 | A | B | C | D | E | F |
\end{Verbatim}

Because of the full bit at level 1 rather than traversing deeper into
the sub block A-B we consider the access to location B to be
\emph{adjacent} (enough) to merit another doubling of the backed
block.

\subsection{Fixed Size Allocations using Ledging}

Doubling is a useful technique to deal with grow able spaces that
cannot be a--priori bounded. For fixed size buffers another population
strategy is called for. In particular we may apply ledging also in the
virtual memory case.

As an example of this consider an application requesting a virtual
space of exactly 11 bytes. Since it is a fixed size buffer we can
pre--populate this space as follows:
\begin{Verbatim}
 Level 4:     .---------------------------------------------------------------.
              :                                                               :
 Level 3:     .-------------------------------.-------------------------------.
              :                               :                               :
 Level 2:     .                               .---------------.- - - - - - - -.
              :                               :               :               :
 Level 1:     .                               .-------.-------.               .
              :                               :       :       :               :
 Level 0:     .                               .       .---.- -.               .
              :                               :       :   :   :               :
              |bbbbbbbbbbbbbbbbbbbbbbbbbbbbbbb|bbbbbbb|bbb|---|---------------|
              | 0 | 1 | 2 | 3 | 4 | 5 | 6 | 7 | 8 | 9 | A | B | C | D | E | F |
\end{Verbatim}
Note that, in this way, we can populate an arbitrary fixed size space
with only logarithmic overhead.

For a fixed size space, writing beyond the last address should trigger
a trap instead of allocating more blocks to back such an
out--of--bounds access. This prevents all sorts of problems with
buffer overflow.

\section{Implementation in Hardware}

It is possible to implement both real memory allocation as well as
virtual memory mapping in hardware. One way to do this would be to
take a dedicated processor and have it run all the code necessary to do
all the tree operations and bookkeeping on both the real memory
allocation tree (rtree) as well as the virtual memory mapping tree
(vtree). This would work although it would incur rather high latency
and achieve only low throughput in terms of the number of
allocations/deallocations that it can process per time unit.

A more efficient design would be to pipeline the operations on both
the rtree as well as the vtree. Note that the pipeline of the vtree
then depends on the pipeline of the rtree as the vtree sometimes needs
to allocate real memory blocks for backing virtual address blocks.

The resulting high level architectural diagram for the rtree would
then look as follows:
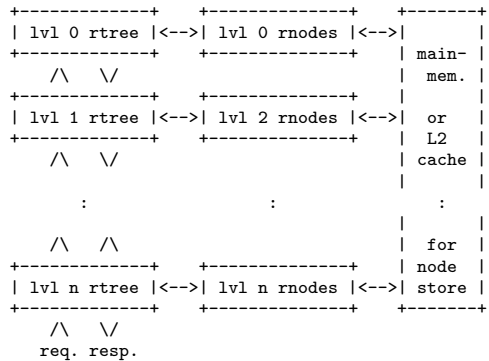
\begin{Verbatim}
             +-------------+    +--------------+    +-------+
             | lvl 0 rtree |<-->| lvl 0 rnodes |<-->|       |
             +-------------+    +--------------+    | main- |
                 /\   \/                            |  mem. |
             +-------------+    +--------------+    |       |
             | lvl 1 rtree |<-->| lvl 2 rnodes |<-->|  or   |
             +-------------+    +--------------+    |  L2   |
                 /\   \/                            | cache |
                                                    |       |
                    :                  :                :
                                                    |       |
                 /\   /\                            |  for  |
             +-------------+    +--------------+    | node  |
             | lvl n rtree |<-->| lvl n rnodes |<-->| store |
             +-------------+    +--------------+    +-------+
                 /\   \/
                req. resp.                                      
\end{Verbatim}
All operations on the rtree can be formulated inductively with respect
to the height of the tree. In addition the decision logic for level
$m+1$ refers only to nodes on level $m+1$ or on level $m$. When care
is taken to separate the node storeage per level it therefore becomes
possible to pipeline mutations on the tree state.

Care should be taken to maintain the invariant that the niche maps
always represent a valid under approximation of the free number of
niches. We propose a reservation system where requests reserve niches
as they descend into the pipeline, and update the niche maps and lift
their reservations as they come out of the pipeline. This way,
requests are never denied because the niche maps led them to a niche
that is, in fact, already taken by a request that just pre--empted it.

The types of request that we can push into the pipeline are:
\begin{enumerate}
  \item request: allocate a block of level $l$, response: the base
    pointer of the allocated block $p$
  \item request: de-allocate the block with base address $p$,
    response: none
\end{enumerate}

In order the improve the independence and optimize the storage for
each level of the pipeline we may give each level its own independent
node store, or node cache.
Note that the operations of level $m+1 \le n$ may involve also nodes
from level $m$ (to reduce clutter these links are not shown in the
diagram). The important thing to note is that the decision logic on
level $m+1$ does \emph{not} require access to levels below $m$ or
above $m+1$.

\newpage

For the vtree a very similar design is possible, although each level
of the vtree pipeline would depend on the rtree pipeline for the
allocation of backing blocks. The resulting high level architectural
diagram for the vtree would look as follows:
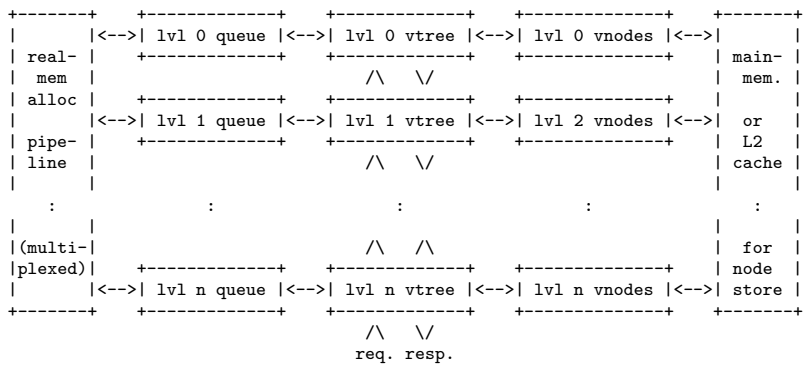
\begin{Verbatim}
+-------+    +-------------+    +-------------+    +--------------+    +-------+
|       |<-->| lvl 0 queue |<-->| lvl 0 vtree |<-->| lvl 0 vnodes |<-->|       |
| real- |    +-------------+    +-------------+    +--------------+    | main- |
|  mem  |                           /\   \/                            |  mem. |
| alloc |    +-------------+    +-------------+    +--------------+    |       |
|       |<-->| lvl 1 queue |<-->| lvl 1 vtree |<-->| lvl 2 vnodes |<-->|  or   |
| pipe- |    +-------------+    +-------------+    +--------------+    |  L2   |
| line  |                           /\   \/                            | cache |
|       |                                                              |       |
    :               :                  :                  :                :
|       |                                                              |       |
|(multi-|                           /\   /\                            |  for  |
|plexed)|    +-------------+    +-------------+    +--------------+    | node  |
|       |<-->| lvl n queue |<-->| lvl n vtree |<-->| lvl n vnodes |<-->| store |
+-------+    +-------------+    +-------------+    +--------------+    +-------+
                                    /\   \/
                                   req. resp.                                      
\end{Verbatim}
Note that by interposing queues between the real memory pipeline and
the vtree pipeline we can prevent (reduce) stalling of the vtree
pipeline waiting on a response of the rtree pipeline by
pre--allocating buffers for the relevant levels.
(At some point the highest levels closest and equal to $n$ should not
be pre--allocated as this would cost too much real memory.)
Amongst the types of requests that we can push into the pipeline are:
\begin{enumerate}
  \item request: create a new virtual space with a certain population
    strategy $s$ and a certain size $x$, response: a logical handle $h$
    to the newly created virtual space
  \item request: destroy virtual space $h$, response: none
  \item request: translate a virtual address $(h, y)$ with handle $h$
    and offset $y$, response: a base pointer $p$ and the level $l$ of
    the enclosing block or a failure code when $y$ is out of bounds.
\end{enumerate}
For the last operation we also obtain the level of the leaf in the
vtree that contained the base pointer $p$. This is useful information
for prefetching neighboring cache lines in order to speed up future
accesses.

\section{Conclusion}\label{sec:conclusion}

  In this report we have discussed the concepts of geometric memory
  alignment, geometric memory allocation and geometric memory mapping.
  We have introduced block trees as an efficient data structure for
  representing geometrically aligned block allocation states.
  We have introduced niche maps as an efficient means to find the right
  place to allocate a chunk of a given size whilst maintaining good
  packing and avoiding fragmentation.
  We have introduced ledging as a process to deal with chunk sizes
  that are not a power of two.
  We have discussed the use of block trees for memory mapping in the context of
  virtual memory management.
  We have shown how ledging can also be applied at the level of
  virtual memory in order to create fixed size virtual memory spaces.
  Finally we have discussed implementation of both geometric memory
  allocation as well as geometric memory mapping in hardware.

In future versions of this report we plan to validate this design
further and add more references to related work and the state of the
art.

\end{document}